\documentclass[12pt]{article}
\usepackage{graphics}
\newcommand{\la}[1]{\label{#1}}

\newcommand{\be}{\begin{equation}}
\newcommand{\ee}{\end{equation}}
\newcommand{\ba}{\begin{eqnarray}}
\newcommand{\ea}{\end{eqnarray}}
\newcommand{\bastar}{\begin{eqnarray*}}
\newcommand{\eastar}{\end{eqnarray*}}

\title{Glueballs, closed fluxtubes and $\eta$(1440) \\ \vskip 0.4cm 
}
\author{ \rm Ludvig\ Faddeev\thanks{Faddeev@euclid.pdmi.ras.ru}, 
\hspace{0.1cm}
\rm Antti \ J. \ Niemi\thanks{Antti.Niemi@teorfys.uu.se} 
\hspace{0.1cm} \rm and \hspace{0.1cm} 
\rm Ulrich \ Wiedner\thanks{Ulrich.Wiedner@tsl.uu.se}
\\ \\ 
$^{*}${\it St. Petersburg Branch of Steklov Mathematical Institute}, \\ 
{\it Russian
Academy of Sciences, Fontanka 27, St.Petersburg, Russia} \\ \\
$^{\dagger}${\it Department of Theoretical Physics, Uppsala University,} \\ 
{\it Box 803, S-75 108 Uppsala, Sweden} \\ \\
$^{\ddagger}${\it Department of Radiation Sciences, Uppsala University,} \\
{\it Box 535,
S-75 121 Uppsala, Sweden} 
}
\begin{document}

%
%
%
\maketitle

\begin{abstract}
\noindent
The $\eta_L(1410)$ component of the $\eta(1440)$ pseudoscalar
has strong affinity for glue. But its mass is incompatible with
lattice simulations that predict a much higher value for the
$0^{-+}$ glueball. Consequently it has been suggested
that $\eta_L(1410)$ signals physics beyond
the Standard Model. Here we argue that if glueballs are 
closed gluonic fluxtubes then $\eta_L(1410)$ is a prime 
candidate for the $0^{-+}$ glueball.
\end{abstract}
\vfill\eject
\noindent
Quantum ChromoDynamics describes the strong interactions of 
quarks and gluons. At high energies QCD becomes asymptotically 
free and explains observed phenomena with an impressive accuracy. 
But at lower energies where the coupling grows and hadrons are 
formed the structure of QCD remains inexplicable. In this regime
strong interactions are ruled by poorly understood non-perturbative 
phenomena such as color confinement and chiral symmetry breaking 
\cite{qftbook}. One of the anticipated predictions 
is the appearance of glueballs as massive and 
color-neutral bound states of massless and confined gluons. 
The lowest mass glueball relates to the first excited state 
in the spectrum of the Yang-Mills theory, thus it is stable within 
the purely gluonic sector of QCD. Lattice computations predict that in 
the limit of infinite quark masses the lowest mass glueball 
is a $J^{PC} = 0^{++}$ state with a mass $1611 \pm 163 \ MeV$
\cite{lattice}. This is within the range of isoscalar $q \bar q$ mesons, 
which can be grouped into SU(3) flavor nonets \cite{pdg} and glueballs 
should appear supernumerous to this nonet.
Some of the additional experimental signatures of a glueball are:   
 
\vskip 0.2cm

{\it i)} Glueballs should be produced in $p \bar p$ annihilation
processes, as a $q \bar q$ annihilation leads to a gluon-rich
environment which strongly favors the formation of gluonic 
degrees of freedom. 

{\it ii)} The central region of various other high-energy hadron-hadron 
scattering processes should similarly favor the production of 
glueballs. 

{\it iii)} The radiative decay of quarkonium states and 
especially the radiative J/${\psi}$ decay should be a prime 
source of glueballs.

{\it iv)} The decay branching fractions of a glueball should be
incompatible with SU(3) predictions for $q \bar q$ states.
 
{\it v)} Since glueballs have no direct coupling to electromagnetism, 
they should be absent in any $\gamma\gamma$ collision processes.

\vskip 0.4cm

Presently, there is a wide consensus that the relatively narrow state
$f_0(1500)$ is the best available candidate for the lowest 
mass $0^{++}$ glueball \cite{pdg}. It does fullfill all of the 
above experimental criteria {\it i)-v)} and its 
mass is well within the range of lattice 
predictions. 

Lattice simulations predict further that the next lowest mass
glueball should be a $2^{++}$ state, with a mass of $2232\pm 
310 \ MeV$ \cite{lattice}. This prediction fits wonderfully 
with the asserted observation of $f_J(2220)$,
which has been proposed as a candidate for the $2^{++}$ glueball. 
Unfortunately, with new data from $p \bar p$ annihilation 
the existence of a $f_J(2220)$ state seems to fade 
away \cite{CB1}. Consequently there does not seem to be any viable candidate 
for the $2^{++}$ glueball with a mass below $ 2.3 \ GeV$,
the regime that has been probed by $p \bar p$ annihilations.

The lattice prediction of the $0^{++} \sim f_0(1500)$ 
glueball is almost perfect. But the difficulties 
in pinpointing a $2^{++}$ glueball, at least at
energies lower than $2.3 \ GeV$ suggests that there might
be place for additional experimental and theoretical input. 
Since glueballs relate directly to confinement and the mass gap in 
Yang-Mills theory, their study is extremely important 
for our understanding of strong interactions and the origin
of mass. Indeed, at the moment our theoretical perception of
glueballs is quite lacking, and rather marred with 
puzzling experimental observations. Notorious in this respect is 
the $J^{PC} = 0^{-+}$ pseudoscalar $\eta(1440)$ which is
most likely a mixture of two particles, the $\eta_H(1480)$ and 
the $\eta_L(1410)$ \cite{pdg}. The $\eta_H(1480)$ couples strongly 
to kaons. In fact, it seems to be an almost ideal mix of $s \bar s$. 
This makes it a likely candidate for the $s \bar s$ member 
of the pseudoscalar nonet. The $\eta_L(1410)$ becomes
then supernumerous, and consequently it is a candidate 
for exotics: In a full analogy with the nearby $f_0(1500)$, 
the $\eta_L(1410)$ is a relatively narrow state that  
fullfills  the above criteria {\it i)-v)} which are 
expected to be satisfied by a glueball,with the sole exception that
for central production it has been seen in high-energetic 
$\pi p$ scattering decaying into $\pi\pi\eta$ \cite{rath} and not 
in the central production of $pp$ scattering. However this 
could simply reflect kinematical suppression.
However, since lattice simulations predict that the 
mass of a $0^{-+}$ glueball should 
be in the vicinity of $2.5 \ GeV$, literature 
\cite{glennys1}, \cite{glennys2} suggests 
that $\eta_L(1410)$ is 
a degree of freedom which is beyond the Standard Model 
QCD, perhaps a bound state of light gluinos. 

Here we shall inspect the possibility that the $\eta_L(1410)$ 
could actually be interpreted as a glueball within standard QCD. 
Specifically, we shall suggest that in the intuitively appealing 
picture where glueballs are viewed as (possibly knotted) 
closed gluonic fluxtubes {\it i.e.} as closed QCD strings, 
the $\eta_L(1410)$ is a natural candidate for the $0^{-+}$ glueball state.  
Furthermore, we shall argue in a quite model independent manner 
that if glueballs are closed fluxtubes there should be a natural
degeneracy between the two glueballs $0^{++} \sim f_0(1500)$ 
and $0^{-+} \sim \eta_L(1410)$. Consequently if
$f_0(1500)$ and $\eta_L(1410)$ are indeed 
glueballs their experimental investigation will directly 
scrutinize physical depictions of strong interactions, 
the string interpretation of confinement and the origin of 
mass in the Universe.    

We shall also propose that a scrutiny of the
slight $90 \ MeV$ mass difference between the $\eta_L(1410)$ 
and the $f_0(1500)$ might have a deep significance:
It could reflect the presence of a new term in
the strong interaction Hamiltonian which breaks exact
mass degeneracy between eigenstates that are related 
to each other by parity. However, we also note that experimentally it is well 
known that interferences with nearby resonances can easily 
shift masses by as much as $5\%$. Consequently the quoted 
values for the $f_0(1500)$, which are in the range of $1445$-$1560 
\ MeV$ dependending on reaction channels, are not necessarily 
inconsistent with an exact mass degeneracy with $\eta_L(1410)$. In fact, 
since $f_0(1500)$ is known to be heavily mixed with nearby meson 
states, a precise mass comparison with $\eta_L(1410)$ becomes 
quite delicate.

Besides (essentially) concurrent masses, relative narrowness, 
and the criteria {\it
i)-v) } which are all satisfied by both $f_0(1500)$ and 
$\eta_L(1410)$ these two states have also various 
additional resemblances. In particular 
their production ratios turn out to be remarkably similar. For example, 
in $p \bar p$ annihilations the production of $f_0(1500)$ has the
following measured branching ratios \cite{CB2},
${\tt BF}[ p \bar p \rightarrow f_0(1500) / 
p \bar p \rightarrow 3\pi^0 ]  \ = \ (13 \pm 4 \%)$ 
and ${\tt BF}[ p \bar p \rightarrow 3\pi^0 ] \ = \ (5.5 \pm 1.0) 
\times 10^{-3}$. When we take into account that $2\pi^0$ decays 
represent $(9.3 \pm 2.5)\%$ of all $f_0(1500)$ decays \cite{AMS98} 
we conclude that 
\be
{\tt BF}[p \bar p \rightarrow f_0(1500)] \ =  \ (7.7 \pm 3.8) \times 
10^{-3}
\la{br1}
\ee

For $\eta_L(1410)$, the $K\bar K\pi$ and $\eta\pi\pi$ decay modes are 
expected to be the dominant \cite{glennys2}. Both have been measured in
$p \bar p$ annihilations, with results \cite{CB4} $ 
{\tt BF} [ p \bar p \rightarrow \eta_L(1410)\pi\pi /$ $
\eta_L(1410) \rightarrow \eta\pi\pi ] \ = \ 
(3.3 \pm 1.0) \times 10^{-3}$ and \cite{bai}
$ {\tt BF} [ p \bar p \rightarrow \eta_L(1410)\pi\pi / 
\eta_L(1410) \rightarrow K\bar K\pi ] \ = \
(2.0 \pm 0.2) \times 10^{-3} $. When we add these, we find 
\be
{\tt BF} [ p \bar p \rightarrow \eta_L(1410)\pi\pi ]
\ = \ (5.3 \pm 1.7) \times 10^{-3}
\la{br2}
\ee
Comparing (\ref{br1}) with (\ref{br2}) and barring for experimental 
uncertainties, we conclude that the $p \bar 
p$ annihilation production rates of the two glueball candidates 
$f_0(1500)$ and $\eta_L(1410)$ are remarkably similar.

We have also compared the observed production rates in 
radiative J/$\psi$ decays. The branching fraction for the production 
of the $f_0(1500)$ and its subsequent decay into $4\pi$ has been 
measured to be $(8.2 \pm 1.7) \times 10^{-4}$ \cite{bug}, with $4\pi$
decays accounting for $(61.7 \pm 9.6)$ per cent of all $f_0(1500)$ decays. 
Therefore, we expect a branching fraction of $ {\tt BF }[ 
J/\psi \rightarrow \gamma f_0(1500)]  \ = \ (1.3 \pm 0.3) 
\times 10^{-3} $ The branching fraction for the production 
of the $\eta_L(1410)$ 
can be determined from its decays into $K\bar K\pi$ 
\cite{Mark1} and $\eta\pi\pi$ \cite{Mark2}. Adding the measured
results leads to the branching fraction $
{\tt BF} [J/\psi \rightarrow \gamma \eta_L(1410)] \ = \ 
(1.0 \pm 0.46) \times 10^{-3}$. 
Again, within experimental uncertainties the radiative 
J/$\psi$ decay production rates of $f_0(1500)$ and
$\eta_L(1410)$ are remarkably similar. 

We find that these similarities in the production rates together
with the relative narrowness of both states and the fact that 
both satisfy all of the overall criteria $i)-v)$, is quite remarkable. 
Indeed, for us this suggests that the natural interpretation of 
$\eta_L(1410)$ is in terms of a $0^{-+}$ glueball, related by
parity to the $0^{++}$ glueball $f_0(1500)$.

We are certainly aware that our proposal is somewhat
unorthodox, and contradicts results from
(quenched) lattice simulations. Consequently there is a 
need for some theoretical backing. For this we proceed
to scrutinize whether our two glueball candidates  
could be interpreted physically as 
closed gluonic fluxtubes {\it i.e.} closed QCD strings. 

The formation of a confining gluonic fluxtube (an open QCD string) between 
two widely separated quarks is widely accepted, and is also 
supported by lattice simulations. From this point of view it becomes 
natural to relish the interpretation that glueballs are indeed closed 
QCD strings, that can be emitted by a (relatively) long linear 
string which connects two widely separated quarks. But a 
straight, linear string has an 
energy which is proportional to its length. 
This implies that in the absense of other contributions to its
energy, a closed string becomes unstable since it can
shrink away by minimizing its length. This instability is present 
for example in the abelian Higgs model with a single complex field; 
it is well known that a closed toroidal vortex ring in a type-II 
superconductor is unstable and shrinks away \cite{ahm}.
If glueballs are closed strings, they must be stable 
against shrinkage within the purely gluonic part of QCD. 
This is a consequence of mass gap and color confinement, 
which prevent the glueballs from decaying into massless gluons.
Besides the linear string tension
there must then be additional contributions to the energy of a closed 
gluonic string. Lattice simulations indicate that a straight
linear gluonic string is only subject to a (classically) linear tension 
\cite{luscher}. Thus any additional force which could stabilize 
a closed toroidal string against shrinkage should have 
a geometric origin, present in a toroidal configuration but 
absent when the string is straight. The natural source of this 
force is in the extrinsic three dimensional geometry of the string, 
in the way how the string twists and bends \cite{nasir}. 
Indeed, suppose that we bend a finite length linear string into 
a toroidal ring. If we twist the string once around its core
before joining its ends to form the ring, 
this can lead to a twisting contribution to the energy which 
prevents the ring from shrinking.

The twisting of a toroidal closed string reveals a natural 
two-fold symmetry in the spectrum: The twist can be either a 
left-handed ($L$) or a right-handed ($R$) rotation around the 
core. We then have two different but stable configurations, 
a left-twisted and the right-twisted closed string. If a Yang-Mills 
theory can indeed differentiate between the left-twisting
and the right-twisting, a parity-invariant 
quantum Yang-Mills Hamiltonian has then a two-fold denegeracy 
among its stringlike eigenstates, $|L>$ and $|R>$, corresponding 
to the left-handed and right-handed twisting along the closed 
gluonic string. The parity operator $\tt P$ that commutes 
with the Hamiltonian relates these states by mapping ${\tt P}|L> = 
|R>$ and ${\tt P}|R> = |L>$. Consequently we can also diagonalize 
${\tt P}$ by setting $ |\pm> \ = \ 1/\sqrt{2}( |L> \pm |R> )$
where $|\pm>$ are parity-even and parity-odd eigenstates of 
the Hamiltonian, ${\tt P}|\pm> = \pm |\pm>$. 
If glueballs correspond to such twisted toroidal
states, in a parity-invariant QCD the glueball spectrum must 
then reflect this degeneracy. In particular, in a 
parity invariant QCD we expect the $0^{++} \sim f_0(1500)$ 
glueball state to naturally relate to a $0^{-+}$ glueball state, 
with $\eta_L(1410)$ a wonderful, experimentally observed 
candidate. 

But we emphasize that parity invariance does not 
necessarily imply equality of masses, the QCD Hamiltonian 
can contain parity invariant terms which remove an exact 
mass degeneracy. For example with ${\tt P}$ a hermitean 
parity operator a perturbation $H \to H + \epsilon {\tt P}$
leads to a mass difference $\Delta m = 2\epsilon$ between 
the parity related $|\pm>$ states. As a consequence 
a small observed mass difference between $f_0(1500)$ 
and $\eta_L(1410)$ could mean that an exact parity degeneracy 
between the Yang-Mills string degrees of freedom 
become broken in QCD.
 
To complete our argument we need to explain how the 
twisting-degree of freedom can be realized in a pure Yang-Mills
theory. Indeed, it is not entirely unnatural to expect 
a $L$-$R$ symmetry to be present, in some form: In the high energy limit 
QCD involves massless gluons that can be prepared in two 
polarization states which can be chosen to
be either left-handed or right-handed. Consequently one can
expect that some kind of $L$-$R$ symmetry could also be
present in the
low energy theory, such as a $L$-$R$ twisting symmetry 
of the closed string. Indeed, this twisting degree of 
freedom can be identified
in the Yang-Mills theory, by employing
an appropriate decomposition of the gauge field $A^a_\mu$ 
\cite{fad1}, \cite{fad2}. 
For notational simplicity but without any loss of generality \cite{bol1}
we consider the decomposition of $A^a_\mu$ in a SU(2) Yang-Mills
theory. Following \cite{fad1} we interpret the Cartan component
$A^3_\mu$ as a U(1)$\ \in \ $SU(2) gauge field. The $A^+_\mu = A^1_\mu +
i A^2_\mu$ together with its complex conjugate then transform
as charged vector fields under the ensuing diagonal SU(2) gauge 
rotations. The two vectors $A^1_\mu$ and $A^2_\mu$ lie in
a plane of a four dimensional space. This plane can be
parametrized by a {\it twobein} ${e^a}_\mu$ ($a=1,2$) with
${e^a}_\mu {e^b}_\nu = \delta^{ab}$. With ${\bf e}_\mu =
1/\sqrt{2}({e^1}_\mu + i {e^2}_\mu)$ we can then represent
the most general $A^+_\mu$ as $ A^1_\mu + iA^2_\mu \ 
= \ i \psi_1 {\bf e}_\mu + i \psi_2
{\bf e}^\star_\mu $ where $\psi_1$ and $\psi_2$ are two complex fields; see 
\cite{fad1} for details. We set $\rho^2 = |\psi_1|^2 + 
|\psi_2|^2$ and define the three-component unit vector
$\vec n$ by
\be
\vec n \ = \ \frac{1}{\rho^2} ( \psi_1^\star \ \psi_2^\star)
\vec \sigma \left( \matrix{ \psi_1 \cr \psi_2 } \right)
\la{defn}
\ee
where $\vec \sigma$ are the standard Pauli matrices.
We substitute the decomposed fields in the Yang-Mills
Lagrangian, and keep {\it only} terms which involve $\rho$
and $\vec n$. The result is \cite{fad1}
\be
- \frac{1}{4} F_{\mu\nu}^2
= (\partial_\mu \rho)^2 + 
\rho^2 (\partial_\mu n^a)^2 + \frac{1}{4} ( \epsilon_{abc}
n^a \partial_\mu n^b \partial_\nu n^c)^2 + V(\rho^2, \vec h \cdot
\vec n) + \dots 
\la{act}
\ee
Here $V$ is a potential term. It involves some of the 
additional terms that we have deleted. This potential
term leads to the breaking of the global $O(3)$ invariance 
of the action under rigid rotations of $\vec n$, giving a
mass to the two Goldstone bosons \cite{fad1}. The 
function $\rho$ relates to the average density of the
two scalars $\psi_1$ and $\psi_2$,
\be
\int (A^1_\mu A^1_\mu + A^2_\mu A^2_\mu) 
=  \int (|\psi_1|^2 + |\psi_2|^2) \ = \ \int \rho^2
\la{zakh}
\ee
The minimization of (\ref{zakh}) along gauge orbits 
selects the background gauge condition $D^{ab}_\mu(A^3) 
A^b_\mu = 0$ {\it w.r.t.} the Cartan $A^3_\mu$ and
in particular the minimum $\rho_{min}$ of (\ref{zakh}) is 
gauge invariant \cite{zakh}. A one-loop computation suggests that 
the average $<\! \rho_{min} \! >$ is nonvanishing \cite{lisa}.
The unit vector $\vec n$ can detect whether
a gluonic fluxtube is left-handed or right-handed. In fact,
when $<\rho_{min}^2> \not= 0$ the action (\ref{act}) is known to 
support stable knotted solitons \cite{nature}, \cite{hieta}
and the simplest soliton  
describes either a left-handed or a right-handed 
unknot {\it i.e.} it is a twisted, $L$-$R$ degenerate 
toroidal state as we desired. Provided these knotted solitons indeed 
survive in the full Yang-Mills quantum theory, we then have 
natural candidates for the glueballs as closed and knotted
$L$-$R$ degenerate fluxtubes.

Obviously it would be too {\it naive} 
to expect the solitons of (\ref{act}) 
to provide a {\it quantitatively} accurate description 
of QCD fluxtubes. For this
we must account for the roughening which is  
due to quantum fluctuations in the additional fields that 
appear in the decomposition of $A^a_\mu$ in the 
full SU(3) Yang-Mills theory. However, it is interesting 
to consider the predictions of these solitons, maybe some of 
their properties are sufficiently universal to survive a more
comprehensive analysis in the full SU(3) theory. 

The energy spectrum $E_Q$ of the solitons in (\ref{act}) follows a 
rational curve in their self-linking number (Hopf invariant)
$Q_H$; see \cite{nature} for details: $E_Q \ \geq \ 
c \cdot |Q_H|^{3/4}$. If $f_0(1500)$ and $\eta_L(1410)$ 
are indeed the lowest mass states, we then have $c 
\approx 1500 \ MeV$ suggesting the 
mass spectrum $M_Q \ \approx \ 1500 \cdot |Q_H|^{3/4} \ 
MeV $. This predicts that the next ($|Q_H|=2$) glueball has
a mass in the vicinity of $2500 - 2600 \ MeV$, well within
the range of the planned radiative charmonium decay 
experiments at CLEO-II and BES. Obviously this estimate 
is very crude, but we note that
there are general topological, model independent
arguments \cite{bol2} which suggest that the 
$3/4$-scaling law should be universal and reliable 
at least for larger values of $|Q_H|$. If this 
persists at lower values of $|Q_H|$, there could be very interesting 
physics around $6.4 - 6.5 \ GeV$ which is near the 
upper reach of energy at the recently approved antiproton 
facility at GSI: This corresponds to $Q_H = \pm 7$, and 
numerical simulations suggest 
that the ensuing soliton is a trefoil which is a nontrivial, 
{\it chiral} knot. 

In three spatial dimensions the self-linking
number of a knot is a topological invariant, 
it remains intact under continuous deformations of the
knot. But if a knot is embedded in a space with more than
three spatial dimensions its self-linking number ceases to be
a topological invariant and the knot can
disentangle. This suggests that the stability and decay
properties of glueballs could be employed to
explore the dimensionality of space-time, how the
strong interaction couples to possible extra dimensions.
In particular, the very presence of a (relatively) stable 
knotted gluonic fluxtube is an indicative that strong
interactions live in three spatial dimensions.

Obviously, a crucial test of our proposal comes 
from comparisons with detailed lattice simulations. 
If correct, the pseudoscalar configurations 
employed untill now in lattice simulations can only 
have a tiny {\it Umklapp} with the lowest mass 
$0^{-+}$ glueball state.

Finally, in full QCD we expect that (virtual) light
$q \bar q$ pairs instabilize a closed and knotted gluonic string: 
A closed string can open itself and become disentangled into another
closed string but with a different self-linking number, through the 
formation and subsequent annihilation of a light quark-antiquark pair. 
This leads to an intuitively very attractive picture of interactions 
between guarks and glueballs, where quarks act much like certain 
enzymes act in the process of DNA replication by allowing one strand 
of the gluonic fluxtube to pass through another, thus changing its 
self-linking number and eventually leading to its decay 
into mesons. In particular, when quark loops are supressed
like in the limit of large quark masses or large-N, we expect the
knotted fluxtubes to become stable provided the space-time is 
four dimensional. 
 
In conclusion, we have inspected the known meson spectrum up to 
energies around $2.3 \ GeV$, which is the upper limit that has been 
experimentally probed by the $p \bar p$ annihilation processes. 
By assuming that glueballs can be viewed as closed gluonic fluxtubes 
we have concluded that the mysterious $\eta_L(1410)$ has a natural 
interpretation as the $0^{-+}$ glueball, parity-related to
the $0^{++}$ glueball $f_0(1500)$. This double-degeneracy of the 
glueball spectrum reflects the left-right twisting symmetry 
of a closed toroidal fluxtube. By employing a decomposed version 
of the gauge field we have also explained how this twisting-degree 
of freedom can be realized in a pure Yang-Mills theory. Our 
arguments are quite general and model-independent, suggesting 
that the interpretation of $\eta_L(1410)$ as the $0^{-+}$ glueball 
provides a test of various qualitative aspects of strong interactions. 
These include the properties of QCD string and confinement, the formation 
of a mass gap, a detailed study of the QCD Hamiltonian, and the
dimensionality of space-time as seen by the strong interaction. 
Our rough estimate of the glueball mass spectrum suggests that all 
these could be studied by the recently approved antiproton facility 
at GSI in Darmstadt.

\vskip 0.2cm
L.F. thanks Yu.A. Simonov for a discussion. A.N. 
thanks A. Guth, J. Kuti, C. Morningstar, M. Shifman and K. 
Zarembo for discussions, and F. Close and G. Farrar for 
communications. U.W. thanks G. Bali for communications. This
work has been supported by fund RFBR 02-01-00085, STINT IG2001-062,
and by VR. This paper is dedicated to the memory of a late colleaque
(U.W.) Lucien Montanet, who always believed that the $\eta(1440)$ 
is something special.

\end{document}